# APPLICATION OF FUZZY LOGIC IN DESIGN OF SMART WASHING MACHINE

Rao Farhat Masood

National University of Sciences and Technology, Pakistan

farhatmasood@outlook.com

## ABSTRACT

*Washing machine is of great domestic necessity as it frees us from the burden of washing our clothes and saves ample of our time. This paper will cover the aspect of designing and developing of Fuzzy Logic based, Smart Washing Machine. The regular washing machine (timer based) makes use of multi-turned timer based start-stop mechanism which is mechanical as is prone to breakage. In addition to its starting and stopping issues, the mechanical timers are not efficient with respect of maintenance and electricity usage. Recent developments have shown that merger of digital electronics in optimal functionality of this machine is possible and nowadays in practice. A number of international renowned companies have developed the machine with the introduction of smart artificial intelligence. Such a machine makes use of sensors and smartly calculates the amount of run-time (washing time) for the main machine motor. Real-time calculations and processes are also catered in optimizing the run-time of the machine. The obvious result is smart time management, better economy of electricity and efficiency of work. This paper deals with the indigenization of FLC (Fuzzy Logic Controller) based Washing Machine, which is capable of automating the inputs and getting the desired output (wash-time).*

## 1. INTRODUCTION

### 1.1. Fuzzy Logic

Fuzzy Logic is the mathematical evaluation of any problem based on the degree of truth. The conventional logic dictates the problem statement be either true or false in entirety. Whereas, fuzzy logic gives the shades of truth and false in the same statement. [1] For example, we have three individuals A, B and C aged, 10, 40 and 70 years old and we ask an individual (D) of 90 years about the old age or youth of A, B and C. D would reply all are young. But if we ask another person of 30 years old, the reply would be B and C are old while A being young. For a 6 feet tall person, 5 feet and 8 inches may not be tall height, but for 5 feet tall person, the same height is considered tall.

### 1.2. Fuzzy Logic vis-à-vis Conventional Logic

The conventional logic doesn't add the perspective in determining the solution to any problem statement, however, in fuzzy logic system, the perspective of observer/evaluator is integrated in offering/suggesting solution to a particular problem. For example, if we tend to control an air conditioner based on the conventional/classical control methods, we would only switch the timer on/off based on the thermostat readings (keeping only one variable in consideration). However, when the same solution is proposed by Fuzzy Logic Model, the timer would be on and off at the same time based on the thermostat reading with the certain degree of truth in both the outputs. Both of the outputs will fire simultaneously but their extent of execution would be different. A Fuzzified Air Conditioner desired to maintain the room temperature at 26 degree Celsius would turn on to a certain degree when the temperature goes beyond 26 and similarly would turn off when the temperature becomes less than 26. [2] Thus in fuzzy logic, we tend to combine the variable statements based on degree of truth or false and then analyze the final statement. In the example of air conditioner, we have kept only temperature variable to simplify the application of fuzzy logic other variable maybe humidity, desired air-flow, moisture etc.

### 1.3. Smart Washing Machine

Consider an individual doing the laundry for the first time and is not familiar with the type of fabric or detergent to be used. In order to solve the laundry problem, he would seek input from professionals to save the effort. [3] Consider a solution combining the inputs from detergent maker, fabric maker and professional laundry servicemen into an implementable model that would facilitate the unaware person doing laundry for the first time. This paper is an effort to propose the same by suggesting a model that can effectively reduce the problem of laundry.

## 2. FUZZY LOGIC CONTROLLER (FLC) FOR SMART WASHING MACHINE (SWM)

### 2.1. Concept

The primary concept in designing a Fuzzy Logic Controller (FLC) [4] lies behind the information gathered from various sources of experience or experts (Laundry-Load Characteristics). For instance, when considering the design of washing machine [3], we have to keep the following information summarized up: -

#### 2.1.1. Type of Clothes (Fabric or Textile)

The nature of clothes that require a wash in the washing machine. The input from textile engineer or a person who deals in fabrics will be desirable.

#### 2.1.2. Type of Detergent (Chemical Property)

Different detergent reacts differently on varied nature of clothes. A chemical engineer who is involved in the manufacture of type of detergent will provide the requisite information.

#### 2.1.3. Type of Stain (Chemical Property)

The chemical property of stain on the piece of cloth will vary. Therefore, in order to remove the stains one has to use different methods.

#### 2.1.4. Temperature of Washing Water

The information regarding the upper and lower threshold of temperature that the fabric of clothes, detergent and stain require.

#### 2.1.5. Other Factors

Among the miscellaneous factors, weight of clothes and water, amount of water soaked after wash-water drainage, spin time required to remove the soaked water and heat required to dry the wash-load.

### 2.2. Design

#### 2.2.1. Parameters

All of the parameters could be gathered in an arduous way and in a smart way as well. Obviously, automating the (laundry load) inputs and making them decide the desired and required (washing parameters) output would be a smart way of doing laundry. [5] Controlling these parameters could lead to a cleaner laundry; conserve water, save detergent, electricity, time, and money.

**2.2.1.1. Inputs**

The characteristics of the laundry load (inputs) include: the actual weight, fabric types, and amount of dirt.

**2.2.1.2. Outputs**

The washing parameters (outputs) include: amount of detergent, washing time, agitation, water level, and temperature.

**2.2.2. Practical Model**

In the practical model implementation, consider, for simplicity, as shown in Figure 1, a washing machine with two inputs (Saturation Time and Dirtiness) and one output (Wash Time). For practical purposes type and amount of detergent has been kept manual. [6]

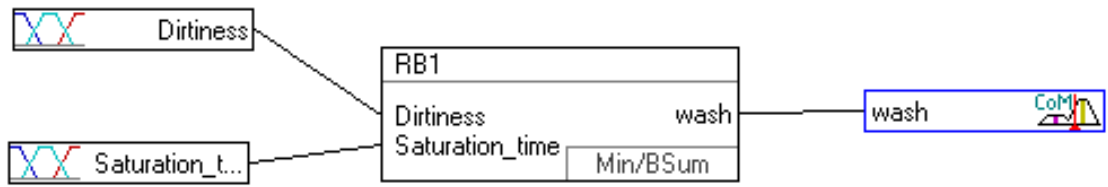

Figure 1. Fuzzy Logic Controller (FLC) of Smart Washing Machine (SWM)

**2.2.2.1. Saturation Time**

The saturation time of dirt or stain in water or conversely the amount of time water takes to saturate with the dirt. It is measured by the opaqueness or opacity of wash water with the use of optical sensor. The output of optical sensor is scaled on fuzzy rule base.

**2.2.2.2. Dirtiness**

It is the type of stain or type of dirtiness of cloth. In order to measure the dirtiness, pressure sensor system is incorporated. It allows the cloth to run a wash-cycle under idle phase and measure the pressure induced by the clothes with the dirtiness. The pressure sensor system is scaled as per fuzzy rule base.

**2.2.2.3. Wash Time**

The output desired or the wash-time required to clean the dirt. It also includes the re-run cycle, stained /saturated wash-water drainage and fresh wash-water refill.

**2.2.3. Fuzzy Subsets [6]**

In practical observation, simple mathematics cannot relate or extract the inputs (Saturation Time and Dirtiness) into meaningful output (Wash-Time). Only a well-informed and knowledgeable user can do the task manually having all the requisite information of clothes material, detergents and nature of wash. With the use of Fuzzy Logic, a rule-base is created based on the knowledge of the user to control or automate the process.

**2.2.3.1. Dirtiness**

The dirtiness is defined in the range from 0 to 30, by defined fuzzy subsets: Low, Medium and High as shown in Figure 2. For ideal demonstration, the subsets are kept triangular, however they can be adoptive in shapes and ranges in reality. At any given instance, all the three subsets will trigger a certain amount of scale. For example, at range 15(exactly middle), High is 0.0,

Low is also 0.0 but Medium subset is 0.99, which means all three subsets will execute in results extraction to 0.0, 0.99 and 0.0 levels.

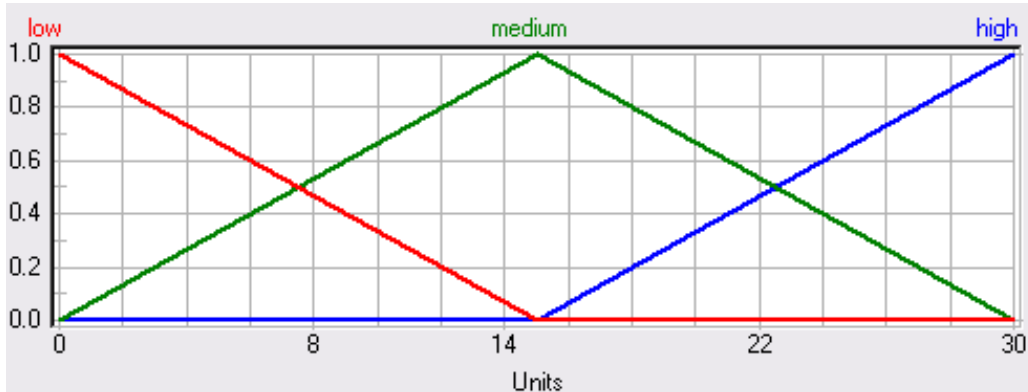

Figure 2. Variable Subset (Dirtiness)

### 2.2.3.2. Saturation Time

The Saturation time is defined in the range of 0 to 10 minutes and has been classified into three sub levels (Low, Medium and High). The same is shown in Figure 3.

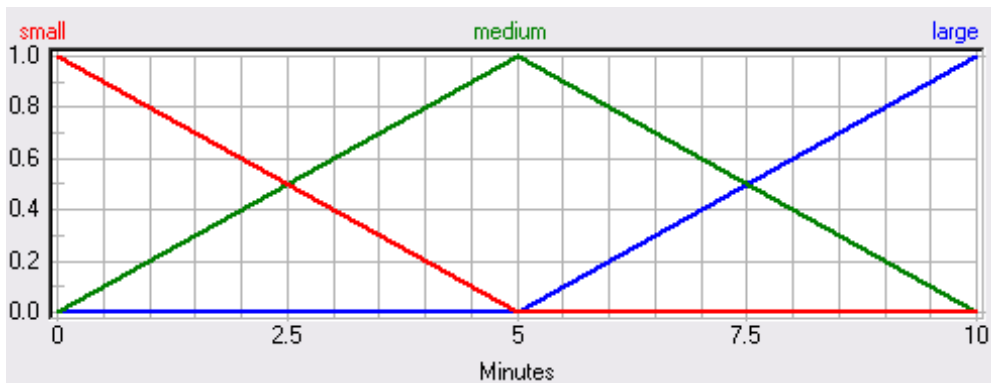

Figure 3. Variable Subset (Saturation Time)

### 2.2.3.3. Wash Time

The wash time is defined in the range from 0 to 15 minutes, based on the fuzzification of above mentioned inputs (saturation time and dirtiness). Further, scaling has been done into five sub levels. The output is defined by the Fuzzy subsets as shown in Figure 4.

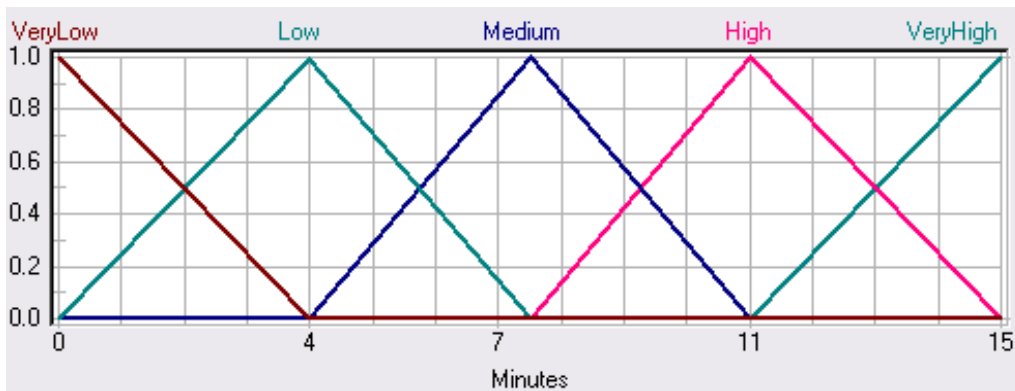

Figure 4. Output Subset (Wash-Time)

### 2.2.4. Output Evaluation

### 2.2.4.1. Rule Base

After fuzzification, the fuzzified input parameters are evaluated using IF/THEN rules. [7] Two control inputs each with 3 fuzzy regions are used in this case. This gives a possible of 3 x 3 = 9 possible input conditions, but for practical purposes in this washing machine only 5 sub levels have been used. However, the algorithm is adaptive and maybe tailored based on more precise input sets. For instance, if DIRTINESS is LOW and SATURATION_TIME is MEDIUM then WASH_TIME is LOW. The complete input and output parameters are given in the table below.

Table 1. Fuzzy Logic Ranges for Output (Wash-Time)

| | | Dirtiness | | |
|---|---|---|---|---|
| | | LOW | MEDIUM | HIGH |
| Saturation Time | SMALL | VERY_LOW | LOW | MEDIUM |
| | MEDIUM | LOW | MEDIUM | HIGH |
| | LARGE | MEDIUM | HIGH | VERY_HIGH |

### 2.2.4.2. Output Sub Levels

The wash time sub levels have been defined in the range given in table 2. For example, if sub level, medium is triggered in output wash-time, the washing time will be from 4 – 11 minutes.

Table 2. Physical Output Ranges (Wash-Time)

| WASH_TIME | |
|---|---|
| LINGUISTIC VARIABLE | RANGE (MINUTES) |
| VERY_LOW | 0-4 |
| LOW | 0-8 |
| MEDIUM | 4-11 |
| HIGH | 7-15 |
| VERY_HIGH | 11-15 |

### 2.2.4.3. Rule Evaluation [6]

Since, we have established that both inputs will trigger a particular sub level of output subset. However, the same is quantified when executed through software. Figure 5 shows that input of dirtiness of medium level is 0.48 of the scale 1 and saturation time of medium level is 0.57. The output is evaluated by ANDing the inputs and resultantly 0.48 of High level of wash time is triggered. And based on the time scales given in table 2, the physical time will be 11 minutes which is approximately half of High sublevel of wash time.

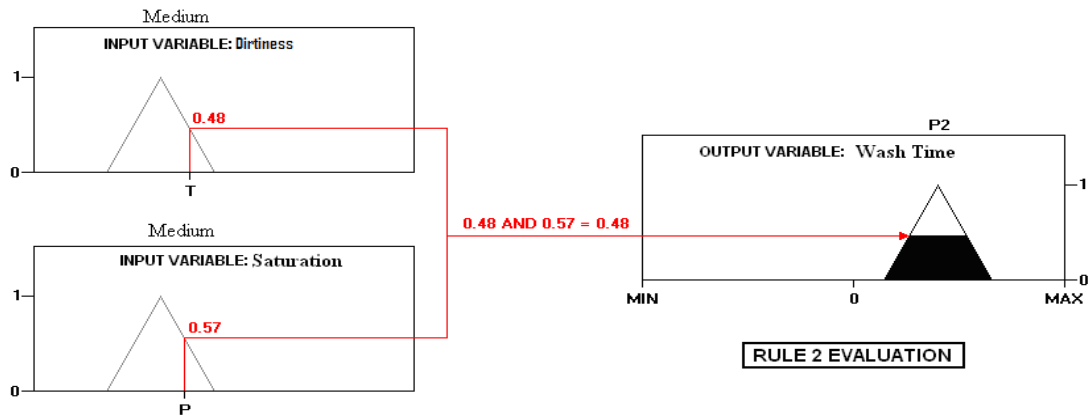

Figure 3. Rule 2 Evaluation

**2.2.4.4. Precision**

For more precise output and inputs, rule base can be modified by defining adaptive subsets and reshaping the sub levels within the subsets [6]. More logic operations can also be introduced for better outcome. The software based precise model having the same number of inputs and wash-time being the realistic output is shown in 3D representation of FLC in figure 6. Here it can be seen that wash-time is continuously varying being non-uniform sub levels of output subset.

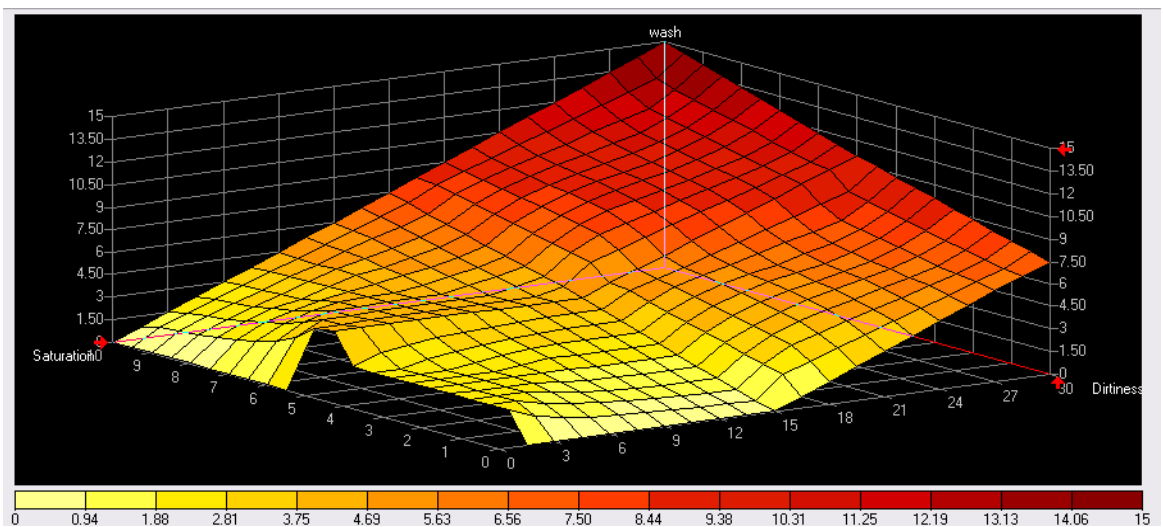

Figure 4. Fuzzy Logic System - 3D Representation

**2.2.4.5. Software**

The figures in this paper were generated using FuzzyTech Designer Software. For the controller part and program analyzing FuzzyTech MCU-51 Edition was used, which supports all 8051 and 80C51 Microcontrollers.

## 3. CONCLUSION

The commercial applications of fuzzy has made its mark over the past few decades because of the theoretically infinite range of control over a particular application. In this paper, the precision in wash time will not only economize energy resources (including electricity & water) but also benefit the user to save finances in commercial laundry solutions offered in the market.

The areas of application of fuzzy logic controllers have more dynamic range as compared to the conventional PID controller. However, there in an importance to highlight that one particular fuzzified solution may not be applicable to all users so there is a dire need to standardize the domain of fuzzified solution.